\documentclass[
 reprint,
 amsmath,amssymb,
 aps,prl,
]{revtex4-2}

\usepackage{appendix}
\usepackage{graphicx}% Include figure files
\usepackage{dcolumn}% Align table columns on decimal point
\usepackage{bm}% bold math
\usepackage[hidelinks]{hyperref}
\usepackage[capitalise]{cleveref}
\usepackage{lipsum}
\usepackage[dvipsnames]{xcolor}
\usepackage[bbgreekl]{mathbbol}
\usepackage{amsfonts}

\DeclareSymbolFontAlphabet{\mathbb}{AMSb}
\DeclareSymbolFontAlphabet{\mathbbl}{bbold}

\usepackage{yfonts}
\DeclareFontFamily{U}{futm}{}
\DeclareFontShape{U}{futm}{m}{n}{
  <-> s * [.97534] fourier-bb % but changing the magnification factor
  }{}
\DeclareMathAlphabet{\mathbbs}{U}{futm}{m}{n}

\begin{document}

\preprint{APS/123-QED}

\title{Universal Response Inequalities Beyond Steady States via Trajectory Information Geometry}

\author{Jiming Zheng}
\email{jiming@unc.edu}
\affiliation{Department of Chemistry, University of North Carolina-Chapel Hill, NC}
\author{Zhiyue Lu}
\email{zhiyuelu@unc.edu}
\affiliation{Department of Chemistry, University of North Carolina-Chapel Hill, NC}

\date{\today}

\begin{abstract}
Fluctuation-dissipation relations elucidate the response of near-equilibrium systems to environmental changes, with recent advances extending response theory to non-equilibrium steady states. However, a general response theory for systems evolving far from steady states has remained elusive. This letter presents a complete trajectory information geometric framework that generalizes response theory for non-stationary Markov processes. By constructing the full trajectory probability manifold and identifying a globally orthogonal coordinate system defined by transition rates, we derive a diagonal Fisher information metric that enables explicit calculations in this high-dimensional space. From the local metric structure, we obtain a Cramér–Rao–type inequality that bounds the linear response of arbitrary non-stationary observables. Furthermore, by analyzing the global geometry of this manifold, we derive a universal non-perturbative (nonlinear) response inequality in terms of geodesic length. This geometric framework reveals deep connections between dynamical activity, observable variance, and system sensitivity, and it encompasses or anticipates several recent results as special cases. Our approach offers new design principles for responsive behaviors in far-from-equilibrium systems.
%Fluctuation-dissipation relations elucidate the response of near-equilibrium systems to environmental changes, with recent advances extending response theory to non-equilibrium steady states. However, a general response theory for systems evolving far from steady states has remained elusive. Using information geometry of stochastic trajectory probabilities, we derive universal thermodynamic bounds on both linear and nonlinear responses of Markov systems to environmental changes, applicable across all non-equilibrium regimes. This theory establishes a new paradigm in non-equilibrium statistical mechanics, offering a unified perspective on the responsiveness of non-stationary systems to external control and environmental changes. Applicable to systems ranging from biological sensory processes to engineered responsive materials, our framework paves the way for understanding and designing complex responsiveness in far-from-equilibrium stochastic systems.

\end{abstract}

\maketitle

{\it Introduction.}---Beyond thermal equilibrium, biological and chemical systems demonstrate complex responsiveness to the changes of their surroundings. Unraveling the principles governing the sensitivity and robustness of systems toward external perturbations is crucial for understanding and designing life-like behaviors in complex systems. While the linear response theory near equilibrium \cite{kubo1966fluctuation,kubo1957statistical} captures the response of systems toward perturbations near thermal equilibrium.
In the past few decades, there have been attempts to extend the linear response theory beyond thermal equilibrium, including violation of fluctuation-dissipation relation \cite{harada2005equality,maes2006time}, results obtained under Gaussian approximation \cite{sato2003relation}, generalization of Einstein relation \cite{baiesi2011modified}, and many other works \cite{maes2009response,PhysRevLett.133.047401,dechant2020fluctuation,maes2020response,baiesi2009fluctuations}. 
Yet, there is still a gap between exact formal response theories and the tangible dynamical properties of systems. Recently, two groups have made prominent results in deriving intuitive bounds on the non-equilibrium responsiveness of stochastic systems\cite{owen2020universal,aslyamov2024nonequilibrium,aslyamov2024general}. However, these recent intuitive results are only applicable to systems restricted to the non-equilibrium steady states (NESS), which can not capture a system's transient response beyond stationary. 
Today, a general framework for understanding the responsiveness of systems far from NESS remains elusive.

In this work, we present a set of universal response relations, \cref{eq: rate response,eq:xi_i_response,eq: finite bound,eq: finite bound 2} for both linear and nonlinear responses of non-equilibrium systems. Here {\it linear response} refers to the response toward infinitesimally small changes of its surroundings, whereas {\it nonlinear response} refers to system's response toward large-amplitude stimuli. In both cases, our theory is applicable to systems that evolve arbitrarily far from NESS. 
Moreover, this theory explicitly relates a system's responsiveness to its kinetic properties, providing direct insight into how dynamical characteristics govern the sensitivity to environmental conditions. 
This universal theory has the potential to transform our understanding of complex responsiveness of non-equilibrium systems and pave the way for the design of rich response features, such as sensitivity\cite{mora2015physical,bialek2005physical,hartich2016sensory}, adaptation \cite{wark2007sensory,lan2012energy,conti2022nonequilibrium}, and robustness\cite{pittendrigh1954temperature,johnson2021circadian,hogenesch2011understanding,ay2007geometric,fu2024temperature}, into far-from-equilibrium systems.

The key of the new theory is the construction of information geometry \cite{amari2016information,ay2017information,ito2018stochastic,crooks2007measuring} in the space of Markovian trajectory probability distributions. Our approach differs from previous information-geometry thermodynamic theories \cite{ito2018stochastic,crooks2007measuring} by focusing on trajectory probabilities\cite{pagare2024stochastic} instead of state probabilities. This novel perspective enables the theory to describe systems evolving far from steady states, significantly expanding the scope of response theory to non-equilibrium and non-stationary relaxation processes.

\begin{figure}[htbp]
    \centering
    \includegraphics[scale=1]{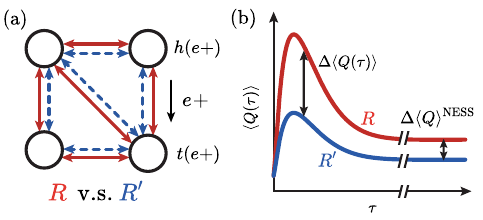}
    \caption{(a) Markov system illustrated by a graph, where the transition of each directed edge (e.g., $e+$) jumps from the head vertex $h(e+)$ to the tail $t(e+)$. (b) System's response to an external change is quantified by comparing the dynamics of the original system $R$ with the changed system $R'$. The transient response is captured by $\Delta \langle Q(\tau)\rangle= | \langle Q[X_\tau]\rangle' - \langle Q[X_\tau]\rangle |$, where $Q(\tau)$ represents a time-dependent observable. As $\tau \to \infty$, both systems approach their respective NESS, and the response converges to $\Delta \langle Q\rangle^{\rm NESS}$. }
    \label{fig: response illustration}
\end{figure}

{\it Setup.}---Consider a Markov process described by a graph $\mathbbl{G} = ( \{\mathbbl{v}_v\}, \{\mathbbl{e}_{e}\})$ with $N_\mathbbl{v}$ vertices $\{\mathbbl{v}_v\}$ and $N_\mathbbl{e}$ undirected edges $\{\mathbbl{e}_e\}$. One can assign each edge $\mathbbl{e}_e$ a forward and a reverse direction, with their corresponding transition probability rates denoted by $(R_{e+}, R_{e-})$. These rates can be alternatively denoted by the two vertices of the edge -- the transition rate from $v'$ to $v$ is $R_{e+}=R_{vv'} $, and equivalently $R_{e-}=R_{v'v}$. Here we call the $v'$ the head of the forward edge: $h(e+)=v'$ and $v$ the tail of the forward edge $t(e+)=v$, as shown in \cref{fig: response illustration}b. 
The dynamics of the Markov system are represented by the master equation
\begin{equation}
    \frac{\mathrm{d}\boldsymbol{p}(t)}{\mathrm{d}t} = R \cdot \boldsymbol{p}(t),
\end{equation}
where $\boldsymbol{p}(t) = ( p_1(t), \cdots, p_{N_{\mathbbl{v}}}(t) )^{\mathsf{T}}$ is the column vector of probability distributions on vertices at time $t$ and $R = \{ R_{vv'} \}_{N_\mathbbl{v}\times N_\mathbbl{v}}$ is the rate matrix with diagonal elements $R_{vv} = -\sum_{v', v'\neq v}R_{v'v}$. 

For a Markov system under the influence of external input or environmental conditions, its transition rates can be dependent on the control parameters (or environmental parameters). To generally describe this dependence, we denote the rate matrix $R (\boldsymbol{\xi})$ as a function of all external control parameters: 
\begin{equation}
    \boldsymbol{\xi} = (\xi_1, \cdots, \xi_{N_p})^{\mathsf{T}}
    \label{eq: xi}
\end{equation}
which is a column vector with $N_p$ control parameters. Examples of environmental parameters $\xi_i$ include but are not limited to temperature, pressure, pH, chemical concentration, electric field, etc. Under this description, alternation in the control parameter $\boldsymbol{\xi}$ changes the system's probability transition rates $R (\boldsymbol{\xi})$, which may result in changes in its dynamics. 

To construct a theoretical framework for system's environmental responsiveness, especially if the system evolve in time rather than maintaining the NESS, we focus on the system's probability distribution stochastic trajectories. The trajectory distribution contains more information \cite{pagare2024stochastic} than the solution of the master equation -- state probability $\boldsymbol{p}(t)$. Here, a trajectory with time length $\tau$ can be denoted by a sequence of $n$ jump events at time $\{t_i\}$:
\begin{equation}
    X_\tau = \left((x_0, t_0), (x_1, t_1), \cdots, (x_n, t_n)\right),
\end{equation}
where $x_i \in \{\mathbbl{v}_v\}$, the initial time is denoted by $t_0 = 0$, and $t_n \le \tau$ is the time when the last jump occurs before $\tau$. 
The probability of the trajectory $X_\tau$ can be represented by the following product,$\mathcal{P}[X_\tau] = \mathcal{P}[X_\tau|x_0] \cdot p_{x_0}(0)$ where $p_{x_0}(0)$ is the initial state probability and $\mathcal{P}[X_\tau|x_0]$ is the conditional path probability of $X_\tau$ given the initial state $x_0$. The conditional path probability can also be represented by \cite{seifert2012stochastic}
\begin{equation}
    \mathcal{P}[X_\tau|x_0] = \prod_{i=1}^{n} R_{x_i x_{i-1}} \prod_{i=0}^n e^{\int_{t_{i}}^{t_{i+1}}R_{x_i x_i}\mathrm{d}t},
\end{equation}
where $t_{n+1} = \tau$ is the time length of the trajectory.

The central result of this letter is the environmental responsiveness of Markov systems without the restriction of NESS. Without losing generality, the environmental change can be arbitrary alternations of $\boldsymbol{\xi}$ and the response can be denoted by the trajectory-averaged changes of arbitrary observable $\langle Q[X_\tau]\rangle$. Here the angular brackets denote an average over the ensemble of trajectories generated from the Markov dynamics $R(\boldsymbol{\xi})$. 
We consider system's response toward two types of environmental changes: (1) {\it perturbative change} or {\it linear response}, $\vert \partial_{\xi_p}\langle Q(\tau)\rangle \vert$, which captures the infinitesimal change of the non-stationary observable caused by the infinitesimal environmental perturbation, and (2) {\it finite change} or {\it nonlinear response}, $\Delta \langle Q(\tau) \rangle=|\langle Q (\tau)\rangle' - \langle Q(\tau)\rangle|$, which quantifies the difference in the system's non-stationary outputs between two finitely different environmental conditions $\boldsymbol{\xi}'$ and $\boldsymbol{\xi}$.

{\it Information Geometry.}---To build a general theory, we construct information geometry \cite{amari2016information,ay2017information} for the space of trajectory probabilities, where the Fisher information metric is chosen as the metric tensor for the parametric probability distributions.

\begin{figure}[htbp]
    \centering
    \includegraphics[scale=0.95]{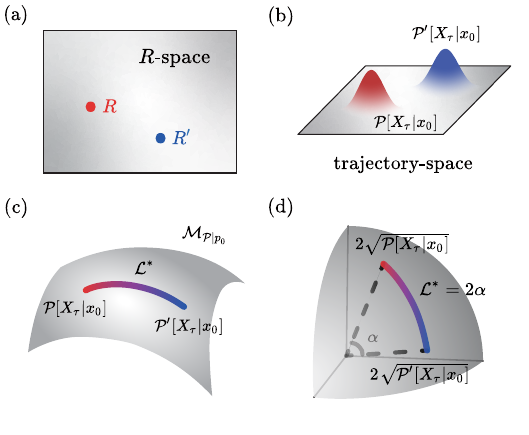}
    \caption{Four equivalent representations of non-stationary Markov Processes. (a) In the space parametrized by rate matrices, the original and changed dynamics are shown by distinct two points. (b) In the trajectory space, the original and changed dynamics ($R$ or $R'$) are represented by their corresponding probability distributions of trajectories conditioned by an initial state distribution $\boldsymbol{p}(0)$. (c) On the manifold of trajectory probability distributions $\mathcal{M}_{\mathcal{P}|p_0}$, the two dynamics are represented by two points $\mathcal{P}[X_\tau|x_0]$ and $\mathcal{P}'[X_\tau|x_0]$. The shortest path between them is the geodesic $\mathcal L^*$. (d) The manifold $\mathcal{M}_{\mathcal{P}|p_0}$ is isometrically embedded to the positive orthant of a radius-$2$ hypersphere of dimension $2N_\mathbbl{e}$, thus the geodesic length between two points becomes an arc length $\mathcal L^*=2\alpha$. }
    \label{fig: manifolds}
\end{figure}

Given a Markov graph and an initial state distribution $\boldsymbol{p}(0)$, the rate matrix $R(\boldsymbol{\xi})$ fully determines the trajectory probability distribution conditioned by the initial $\boldsymbol{p}(0)$. This correspondence between the rate matrix and conditional path probability is shown by \cref{fig: manifolds}(a) and (b). As a result, the manifold $\mathcal{M}_{\mathcal{P}|p_0}$ of conditional path probabilities $\mathcal{P}[X_\tau|x_0]$, as shown by \cref{fig: manifolds}(c), can be fully represented by transition rate matrix $R$. Since the degree of freedom for $R$ is $2N_\mathbbl{e}$, the resulting conditional path probability manifold $\mathcal{M}_{\mathcal{P}|p_0}$ is $2N_\mathbbl{e}$-dimensional. In other words, \cref{fig: manifolds}(a) and (c) share the same dimensionality, and every point on the manifold uniquely specifies an evolution process (i.e., a distribution of trajectories). One can alternatively represent the conditional path probability manifold by embedding it on a hypersphere of radius $2$ \cite{ay2017information}, illustrated by \cref{fig: manifolds}(d). The hypersphere representation provides a geometric argument for the nonlinear responsiveness relation.

As mentioned above, the natural and complete choice of the coordinate system for the $2N_\mathbbl{e}$-dimension manifold $\mathcal{M}_{\mathcal{P}|p_0}$ is the set of all transition rates $\{R_{e+}\} \cup \{R_{e-}\}$. We refer to this choice as {\it transition rate coordinates}. With the help of the trajectory KL divergence recently derived in \cite{pagare2024stochastic}, 
\begin{align}
    D_{\text{KL}} &[\mathcal{P}[X_\tau|x_0]\Vert\mathcal{P}'[X_\tau|x_0]] \nonumber \\
    &= \int_0^\tau \sum_{e\pm} R_{e\pm} p_{h(e\pm)} \left( \ln\frac{R_{e\pm}}{R'_{e\pm}} + \frac{R'_{e\pm}}{R_{e\pm}} - 1 \right) \mathrm{d}t,
\end{align}
whose second-order derivatives (Hessian matrix) gives the {\it geometric foundation} of this work -- the Fisher information metric:
\begin{equation}
    g(\{R_{e+}\}, \{R_{e-}\}) = \text{diag }\left\{ \left\{\frac{\mathcal{A}_{e+}}{R_{e+}^2}\right\}, \left\{\frac{\mathcal{A}_{e-}}{R_{e-}^2}\right\} \right\}.
    \label{eq: FIM directed rate}
\end{equation}
where $\mathcal{A}_{e+} = \int_0^\tau R_{e+}p_{h(e+)}(t)\mathrm{d}t$ denotes the {\it directed activity} on the directed edge $e+$ \footnote{For systems evolving under time-dependent rate matrix, the Fisher information is discussed in the SI.I.}. $\mathcal{A}_{e+}$ is the expectation value of the number of directed transitions along edge $e+$ within the time duration $\tau$ \cite{maes2020frenesy}, which was shown to play an important role in various thermodynamic relations by recent studies \cite{baiesi2009fluctuations,garrahan2017simple,di2018kinetic,shiraishi2018speed,harunari2022learn,hasegawa2024thermodynamic}.

The diagonal form of the Fisher information metric implies that the contribution of response on each edge is additive and statistically independent, as detailed explained in the more recent work \cite{zheng2025spatial}. Geometrically, it also implies that the transition rate coordinate system serves as an orthogonal basis everywhere, making it straightforward to transform to other coordinate systems (e.g., the explicit environmental parameter coordinate system $\boldsymbol{\xi}$). Taking advantage of the orthogonal basis, it is also straightforward to define and evaluate various geometric quantities on the manifold. For example, the line element $\mathrm{d}s^2$ in terms of transition rates is then given by $\mathrm{d}s^2 = \sum_{e\pm} \frac{\mathcal{A}_{e\pm}}{R_{e\pm}^2} \mathrm{d}R_{e\pm}^2$. For an infinitesimal displacement along any axis of the transition rate coordinates, e.g., $R_{e+}$, the line element is simply $\mathrm{d}s^2 = \frac{\mathcal{A}_{e+}}{R_{e+}^2} \mathrm{d}R_{e+}^2$. The geodesic between any two points on the manifold is the path of minimum curve length, illustrated by $\mathcal L^*$ in two representations of the manifold \cref{fig: manifolds}(c) and (d).

Within the general framework of information geometry, this work describes system's non-stationary responses to both perturbative and finite environmental changes, i.e., (1) linear response beyond stationarity and (2) nonlinear response relations.

{\it Linear Response Beyond Stationarity.}---By applying the Cram\'er-Rao inequality \cite{rao1992information,cramer1999mathematical} to the trajectory Fisher information metric defined in \cref{eq: FIM directed rate}, we obtain the {\it first central result} of this work: the thermodynamic bound for the response magnitude of any observable $Q$ to the perturbation. 
The result takes two forms. When considering system's response to the perturbation of a single transition rate, the responsiveness relation can be expressed by:
\begin{equation}
    \left| \frac{\partial \langle Q(\tau) \rangle}{\partial {R_{e\pm}}} \right| \le \sqrt{\operatorname{Var}[Q]} \cdot \frac{\sqrt{\mathcal{A}_{e\pm}}}{R_{e\pm}},
    \label{eq: rate response}
\end{equation}
where the only change is applied to a single directed edge $R_{e+}$ (or $R_{e-}$), while all other transition rates remain intact.
Here, the expectation value of the observable is the ensemble average over all stochastic trajectories of time-length $\tau$: $\langle Q(\tau) \rangle = \int \mathcal{D}[X_{\tau}]  \mathcal{P}[X_{\tau}] Q[X_\tau]$. Under different choices of the exact form for $Q[X_\tau]$, the definition can generally represent either a trajectory observable (time-accumulated quantity) or a transient observable. To realize a transient observable at any time $t$, one can introduce a delta function peaked at time $t$ into $Q[X_\tau]$. Formal discussions on different types of transient observables and their corresponding numerical results can be found in SI.II. If one chooses $Q[X_\tau]$ as a time accumulation observable, our response relations can provide NESS response relations. The linear response inequality \cref{eq: rate response} can also be derived from the Dechant-Sasa response relation \cite{dechant2020fluctuation}. See Sec.VI of SI for the alternative derivation.

In general, \cref{eq: rate response} implies that the responsiveness of an observable is bounded by the product between the square root of observable variance $\operatorname{Var}[Q] \equiv \langle Q(\tau)^2 \rangle - \langle Q(\tau) \rangle^2$ and the square root of the scaled dynamical activity of the perturbed transition rates. Also, in alignment with intuition, this result indicated that the directed edge with larger activities may result in a stronger response toward the perturbation of its rate.

More conveniently, the above response relation can be formulated in terms of environmental-variable responsiveness relation. Through coordinate transformation into the environmental coordinate $\boldsymbol{\xi}$ in \cref{eq: xi}, the responsiveness bound to any chosen environmental variable change $\delta \xi_i$ can be obtained:
\begin{equation}
    \left| \frac{\partial \langle Q \rangle}{\partial \xi_i} \right| \le \sqrt{\operatorname{Var}[Q]} \cdot \sqrt{\sum_{e\pm} \frac{\mathcal{A}_{e\pm}}{R_{e\pm}^2} \left( \frac{\partial R_{e\pm}}{\partial \xi_i} \right)^2},
    \label{eq:xi_i_response}
\end{equation}
where $\frac{\partial R_{e\pm}}{\partial \xi_i}$ denotes the transition rate's dependence on the environmental variable $\xi_i$ while fixing all other environmental conditions. An illustrative example of the application of this relation to illustrate the system's environmental responsiveness can be found in the example at the end of this letter. More formal analysis on the change of coordinate system to study system's responses to dissipative thermodynamic driving forces can also be found in SI.III.

{\it Nonlinear Responsiveness Relations.}---Based on the geometry of the whole manifold $\mathcal{M}_{\mathcal{P}|p_0}$, we can obtain the {\it second central result}, universal responsiveness relations toward finite environmental condition change $\boldsymbol{\xi}$ to $\boldsymbol{\xi}'$. For comparison, the linear-responsiveness relations only utilize the local geometric property of the manifold. 

First, the following {\it geometric non-linear responsiveness relation} holds:
\begin{equation}
    \arctan \frac{|\langle Q(\tau) \rangle' - \langle Q(\tau) \rangle|}{\sqrt{\operatorname{Var}[Q]} + \sqrt{\operatorname{Var}'[Q]}} \le \frac{\mathcal{L}^*(\tau)}{2} \le \frac{\mathcal{L}(\tau)}{2}, \label{eq: finite bound}
\end{equation}
where the $\mathcal{L^*}$ is the geodesic length between the two path distributions for the two environmental conditions $\boldsymbol{\xi}$ and $\boldsymbol{\xi}'$. Geometrically, the ratio $\mathcal{L^*}/2=\alpha$ is the angle shown in \cref{fig: manifolds}(d); and the second inequality in \cref{eq: finite bound} comes from the fact that the geodesic length $\mathcal{L}^*$ is shorter than any other curve length $\mathcal{L}$. Here we define a curve length $\mathcal{L}$ by accumulated difference on all the changed rates:
\begin{align}
\label{eq: L define}
    \mathcal{L}(\tau) = \int_{R}^{R'} \sqrt{ \sum_{e\pm} \frac{\mathcal{A}_{e\pm}}{R^2_{e\pm}} (\mathrm{d}R_{e\pm})^2 }.
\end{align}

By combining \cref{eq: finite bound,eq: L define} and Jensen's inequality \cite{jensen1906fonctions}, one can further obtain an alternative {\it dynamical nonlinear responsiveness relation} that {\it explicitly reveals} the connection between system's responsiveness and its dynamical rate changes:
\begin{align}
\label{eq: finite bound 2}
    \arctan \frac{|\langle Q(\tau) \rangle' - \langle Q(\tau) \rangle|}{\sqrt{\operatorname{Var}[Q]} + \sqrt{\operatorname{Var}'[Q]}}  \le \sqrt{\tau} \sum_{e\pm}\left| \sqrt{R'_{e\pm}} - \sqrt{R_{e\pm}} \right|.
\end{align}

The geometric and dynamical nonlinear responsiveness relations, \cref{eq: finite bound,eq: finite bound 2}, reveal that observables of two different Markov dynamics are bounded by the fluctuations of the observables and a geometric length $\mathcal L{\tau}$. Furthermore, the alternative bound replaced the geometric length by a simple term that scales with the square root of time duration and the square root of the difference in transition rates. Notice that in both inequalities, the right-hand side only involves the transitions whose rate differs for the two Markov systems. Additionally, our result reveals that the difference between the two Markov system dynamics, reflected by an observable $Q[X_{\tau}]$, is geometrically bounded by an angle $\alpha$ (see \cref{fig: manifolds}(d)). This geometric argument is elaborated below.

{\it Geometric Proof of Nonlinear Response Relation.}---Let us consider alternative representations of the space of conditional trajectory probabilities as illustrated by \cref{fig: manifolds}. Notice that each trajectory probability density $\mathcal{P}[X_\tau; R]$ is a map from the product space of the Hilbert space of trajectories $\Omega$ and the $2N_\mathbbl{e}$ dimensional rate parameter space $\Xi$ to a positive real number, i.e., $\mathcal{P}[X_\tau; R]: \Omega \times \Xi \to \mathbb{R}^+$. 
For two different rate matrices $R$ and $R'$, the corresponding trajectory densities are represented by $\mathcal{P} \equiv \mathcal{P}[X_\tau; R]$ and $\mathcal{P}' \equiv \mathcal{P}[X_\tau; R']$. Inspired by reference \cite{ay2017information}, we define their corresponding $\theta${\it -scaled half density} via a map $\pi^{1/2}_\theta: \mathcal{P} \mapsto \rho \equiv \theta \sqrt{\mathcal{P}}$ with a positive real number $\theta$. Here $\pi^{1/2}_\theta$ is a diffeomorphism and preserves the geodesic. 

The map $\pi^{1/2}_\theta$ provides us with an alternative representation of the trajectory probability manifold $\mathcal{M}_{\mathcal{P}|p_0}$ with fisher information metric as a hypersphere in Euclidean space with radius $\theta$. In the half-density representation, the inner product is defined by integrating over the trajectory Hilbert space $\Omega$:
\begin{equation}
    (\rho, \rho') = \int \rho\rho' \mathcal{D}[X_\tau],
\end{equation}
Thus, due to normalization over trajectory space $\Omega$, the self inner product for each $\rho$ must be $(\rho, \rho) \equiv \Vert \rho \Vert^2 = \theta^2$. In other words, all $\rho's$ lives on the positive orthant of the radius-$\theta$ hypersphere of dimension $2N_\mathbbl{e}$ as shown in \cref{fig: manifolds}(d). 

By choosing $\theta=2$, the map $\pi^{1/2}_2$ becomes a isometric diffeomorphism. In other words, the geodesic length on the hypersphere between $\rho$ and $\rho'$ coincides with the geodesic length $\mathcal {L}^*$ between $\mathcal {P}$ and $\mathcal {P}'$. Geometrically, by representing the former as the arc length on the great circle with radius $\theta = 2$, we have $\mathcal {L}^* = 2\alpha$, where $\alpha$ is the geodesic angle as shown in \cref{fig: manifolds}(d).  

Under this geometric analysis, one can further obtain 
\begin{equation}
    \cos\alpha = \frac{(\rho, \rho')}{\Vert\rho\Vert \cdot \Vert\rho'\Vert} = \int_{\Omega} \sqrt{\mathcal{P}}\sqrt{\mathcal{P}'}\mathcal{D}[X_\tau].
\end{equation}
By combining this with the inequality for Hellinger distance from \cite{nishiyama2020tight}, we arrive at 
\begin{equation}
    \cos \alpha \le \left[ \left( \frac{\langle Q(\tau) \rangle' - \langle Q(\tau) \rangle}{\sqrt{\operatorname{Var}'[Q]} + \sqrt{\operatorname{Var}[Q]}} \right)^2 + 1 \right]^{-1/2}.
    \label{SIeq: Bhat bound}
\end{equation}
By rearrangements, we can obtain the non-linear response inequality \cref{eq: finite bound}. More details of the derivation can be found in SI.IV-A.

From this geometry perspective, one can consider the non-perturbative (nonlinear) response relations, \cref{eq: finite bound,eq: finite bound 2}, as extensions of the linear response relation, \cref{eq: rate response}. 
The nonlinear relation reduces to the linear case in the limit of $R' \to R$. In this case, the geodesic shrinks to a point and the geodesic length $\mathcal{L^*}$ reduces to the local Fisher information metric, \cref{eq: FIM directed rate}, and the linear-response relation \cref{eq: rate response,eq:xi_i_response} is recovered.

{\it Discussion: Zoology of Thermodynamic Information Geometry.}---This letter provides a novel trajectory-ensemble information geometric framework with four equivalent representations (shown in \cref{fig: manifolds}) and a complete orthogonal coordinate basis (as shown by \cref{eq: FIM directed rate}). The trajectory-ensemble framework is inherently different from previous state-ensemble information geometric theories 
\cite{crooks2007measuring,sivak2012thermodynamic,ito2018stochastic}, and short-time trajectory theory \cite{ito2024geometric}. Our trajectory-ensemble theory is capable of describing non-equilibrium and non-stationary Markov processes.  

This new framework defines information geometry on the full manifold of the Markov trajectory ensembles, $\mathcal{P}[X_\tau]$, which can be obtained by combining the manifold of conditional distributions of trajectories  $\mathcal{M}_{\mathcal{P}|p_0}$ with the initial state ensemble manifold $\mathcal{M}_{p}$. \footnote{The full manifold is locally the product manifold $\mathcal{M}_p \times \mathcal{M}_{\mathcal{P}|p_0}$}. The metric for the manifold $\mathcal{P}[X_\tau]$ is the product metric $g_p \oplus g_{\mathcal{P}|p_0}$. Here, we focus on the property of the conditional trajectory manifold $\mathcal{M}_{\mathcal{P}|p_0}$.

Our geometric framework offers a complete description of the manifold $\mathcal{M}_{\mathcal{P}|p_0}$. In this framework, it is not necessary to include the trajectory length $\tau$ as a parameter. This is because the time coordinate is dependent on the transition rate coordinates $(\{R_{e+}\} \cup \{R_{e-}\})$. In other words, measuring a system within a longer time period is equivalent to measure a system with faster transition rates, i.e., $\mathcal{P}[X_\tau; \{\theta R_{e\pm}\}] = \mathcal{P}[X_{\theta\tau}; \{R_{e\pm}\}]$ for $\theta \in (0, +\infty)$, as pointed out by the relevant work \cite{di2018kinetic,hasegawa2023unifying}. This time-rescaling equivalence illustrates that the trajectory information geometry in the recent works \cite{hasegawa2024thermodynamic,hasegawa2023unifying} can be reduced from the new complete geometric framework by scaling all the transition rates with the same factor $\theta$, where the length $\mathcal{L}$ is related to $\frac{1}{2}\int_0^\tau\frac{\sqrt{\mathcal{A}}}{t}\mathrm{d}t $. In other words, the recently developed trajectory information geometry analysis \cite{hasegawa2024thermodynamic,hasegawa2023unifying} can be considered as one-dimensional restricted studies of the complete $2N_\mathbbl{e}$-dimensional geometric framework proposed by this Letter. Moreover, one can reduce our results \cref{eq: rate response} to the kinetic uncertainty relation \cite{di2018kinetic} by using the equivalence property.

Here, we conjecture that the responsiveness inequality may be better saturated when the observable $Q$ is strongly correlated with the transition events of the perturbed edge, as illustrated in the SI.VII. More technical discussions of the saturation condition for the inequalities are provided in SI.V.

Notice that the local geometric property of the trajectory manifold, \cref{eq: FIM directed rate}, obtained by the Fisher information of classical Markov trajectories, agrees with previous works on the quantum Fisher information for Lindblad equation \cite{macieszczak2016dynamical,comm}. In the future, it is useful to explore if the information geometric framework developed for classical Markov systems in this work, especially the nonlinear responsiveness relations, \cref{eq: finite bound,eq: finite bound 2}, can be extended by considering quantum information geometry approaches \cite{guta2017information}.

{\it Application: Responsiveness Design Principle.}---Our responsiveness relation provides a general theoretical principle to guide the design of a system's non-stationary response to environmental condition changes. This relation connects experimentally accessible measurements (observable $Q$) with system's inherent transition statistics:
\begin{align}
    \eta^Q_i \equiv \frac{(\partial_{\xi_i}\langle Q \rangle)^2}{\operatorname{Var}[Q]} &\leq \mathcal{A} \cdot \sum_{e\pm} a^2_{e\pm}(\xi_i) \frac{\mathcal{A}_{e\pm}}{\mathcal{A}} \\
    &= \mathcal{A} \cdot \overline{a_i^2} \equiv \eta^\text{Bound}_i,
    \label{eq:stat_rep}
\end{align}
where $\eta^Q_i$ denotes the sensitivity-to-noise ratio for observable $Q$ with respect to control parameter $\xi_i$,  $a_{e\pm}(\xi_i) \equiv \partial_{\xi_i} \ln R_{e\pm}$ is the {\it logarithm sensitivity} of each transition's rate with respect to the control variable $\xi_i$, $\mathcal{A}$ is the total activity, ${\mathcal{A}_{e\pm}}/{\mathcal{A}}$ is the fraction of transitions from each edge, and $\overline{a_i^2}$ is the statistical average of the {\it logarithm rate sensitivity} for all observed transition events. This result aligns with intuition: the parameter sensitivity of a non-equilibrium process depends on the total activity $\mathcal A$ and the average logarithmic rate sensitivity of observed transitions. By further choosing an observational time scale to maintain a constant activity $\mathcal A$, system's sensitivity only depends on the average logarithmic rate sensitivity of observed edges, $\overline{a^2_i}$.

This result can be illustrated with a thermal system described by the generalized Arrhenius law
\begin{equation}
    R_{e\pm} =C_{e} e^{-\beta (B_e-E_{h(e\pm)}) +\beta w_{e\pm}},
\end{equation}
where $\beta=1/k_B T$ is the inverse temperature, $B_e$ denotes energy barrier for edge $e$, $C_e$ and $c_e$ are constants for edge $e$, $E_{h(e\pm)}$ denotes the energy of the head of the directed transition edge $e\pm$, and $w_{e\pm}$ denotes the non-equilibrium work applied to assist transition $e\pm$ that breaks the detailed balance. In this case, the logarithm rate sensitivity for inverse temperature is the generalized activation energy $a_{e\pm} = w_{e\pm}-(B_e-E_{h(e\pm)})$. To design a system with ultra-low (or high) temperature sensitivity, one needs to minimize (or maximize) $\eta_\beta^{\text{Bound}}$ via optimizing the energy landscapes and the external driving patterns. Moreover, to create a system sensitive to one variable ($\xi_i$) but robust to another ($\xi_j$), one can focus on maximizing the ratio between their respective sensitivity bounds $\eta_i^{\text{Bound}}/\eta_j^{\text{Bound}}$.

{\it Conclusion.}---In summary, this work presents a set of universal responsiveness relations for non-stationary Markov processes under either infinitesimal or finite-amplitude perturbations. These relations explicitly reveal the connection between any observable's environmental response, the variance of the measured observable and the dynamical features of the system. It provides a powerful tool for understanding and predicting the environmental responsiveness of complex systems far from equilibrium, with a wide range of potential applications in biological sensory processes and synthetic responsive materials. The illustrative example reveals that the explicit responsiveness relations presented in this work may lead to practical design principles to help the design and optimization of stochastic sensors, biological or synthetic, with desired time-dependent sensitivity or robustness.

{\it Acknowledgments.}---This work is funded by National Science Foundation Grant DMR-2145256. The authors appreciate discussions and suggestions on the manuscript from Prof. Chris Jarzynski and from attendees at the Frontiers of Non-equilibrium Physics 2024 workshop held in July 2024 at Kyoto University, Japan. We also appreciate the careful read and valuable suggestions from the reviewers. 
\bibliography{apssamp}

\end{document}